\documentclass[twocolumn,showpacs,preprintnumbers,amsmath,amssymb,superscriptaddress,floatfix,nofootinbib]{revtex4}

\usepackage{graphicx}
\usepackage{epsfig}
\usepackage{epstopdf}
\usepackage{hyperref}
\usepackage{amsmath}
\usepackage{amsfonts}
\usepackage{amssymb}
\usepackage{mathtools}
\usepackage{upgreek}
\usepackage{bm}

\begin{document}

\title{Theoretical study of the $\Omega(2012)$ state in the $\Omega_c^0 \to \pi^+ \Omega(2012)^- \to \pi^+ (\bar{K}\Xi)^-$ and $\pi^+ (\bar{K}\Xi\pi)^-$ decays}
\date{\today}

\author{Chun-Hua Zeng}
\affiliation{Institute of Modern Physics, Chinese Academy of
Sciences, Lanzhou 730000, China} \affiliation{School of Nuclear
Sciences and Technology, University of Chinese Academy of Sciences,
Beijing 101408, China}

\author{Jun-Xu Lu}
\affiliation{School of Physics $\&$ Beijing Advanced Innovation
Center for Big Data-based Precision Medicine, Beihang University,
Beijing 100191, China}

\author{En Wang} \email{wangen@zzu.edu.cn}
\affiliation{School of Physics and Microelectronics, Zhengzhou
University, Zhengzhou, Henan 450001, China}

\author{Ju-Jun Xie} \email{xiejujun@impcas.ac.cn}
\affiliation{Institute of Modern Physics, Chinese Academy of
Sciences, Lanzhou 730000, China} \affiliation{School of Nuclear
Sciences and Technology, University of Chinese Academy of Sciences,
Beijing 101408, China} \affiliation{School of Physics and
Microelectronics, Zhengzhou University, Zhengzhou, Henan 450001,
China}

\author{Li-Sheng Geng}\email{lisheng.geng@buaa.edu.cn}
\affiliation{School of Physics $\&$ Beijing Advanced Innovation
Center for Big Data-based Precision Medicine, Beihang University,
Beijing 100191, China} \affiliation{School of Physics and
Microelectronics, Zhengzhou University, Zhengzhou, Henan 450001,
China}

\begin{abstract}

We report on a theoretical study of the newly observed
$\Omega(2012)$ resonance in the nonleptonic weak decays of
$\Omega_c^0 \to \pi^+ \bar{K}\Xi^*(1530) (\eta \Omega) \to \pi^+
(\bar{K}\Xi)^-$ and $\pi^+ (\bar{K}\Xi\pi)^-$ via final-state
interactions of the $\bar{K}\Xi^*(1530)$ and $\eta \Omega$ pairs.
The weak interaction part is assumed to be dominated by the charm
quark decay process: $c(ss) \to (s + u + \bar{d})(ss)$, while the
hadronization part takes place between the $sss$ cluster from the
weak decay and a quark-antiquark pair with the quantum numbers
$J^{PC} = 0^{++}$ of the vacuum, produces a pair of
$\bar{K}\Xi^*(1530)$ and $\eta \Omega$. Accordingly, the final
$\bar{K}\Xi^*(1530)$ and $\eta \Omega$ states are in pure isospin $I
= 0$ combinations, and the $\Omega_c^0 \to \pi^+ \bar{K}\Xi^*(1530)
(\eta \Omega) \to \pi^+ (\bar{K}\Xi)^-$ decay is an ideal process to
study the $\Omega(2012)$ resonance. With the final-state interaction
described in the chiral unitary approach, up to an arbitrary
normalization, the invariant mass distributions of the final state
are calculated, assuming that the $\Omega(2012)$ resonance with
spin-parity $J^P = 3/2^-$ is a dynamically generated state from the
coupled channels interactions of the $\bar{K}\Xi^*(1530)$ and $\eta
\Omega$ in $s$-wave and $\bar{K}\Xi$ in $d$-wave. We also calculate
the ratio, $R^{\bar{K}\Xi\pi}_{\bar{K}\Xi} = {\rm Br}[\Omega_c^0 \to
\pi^+ \Omega(2012)^- \to \pi^+ (\bar{K}\Xi \pi)^-] / {\rm
Br}[\Omega_c^0 \to \pi^+ \Omega(2012)^- \to \pi^+ (\bar{K}\Xi)^-$].
The proposed mechanism can provide valuable information on the
nature of the $\Omega(2012)$ and can in principle be tested by
future experiments.

\end{abstract}

\maketitle

\section{Introduction}

The study of baryon spectroscopy is one of the most important issues
in hadron physics and is an essential tool to analyze baryon
structure. Data on baryon masses and decay modes are compiled in the
\textit{Particle Data Book Review}~\cite{Tanabashi:2018oca}. For the
light flavor hyperons with strangeness $-3$, not much is known about
their properties~\cite{Tanabashi:2018oca}. In 2018, the Belle
collaboration reported an $\Omega$ exited state, $\Omega(2012)$, in
the $K^-\Xi^0$ and $K^0_S\Xi^-$ invariant mass
distributions~\cite{Yelton:2018mag}, with the measured mass $M =
2012.4 \pm 0.7 \pm 0.6$ MeV and width $\Gamma = 6.4^{+2.5}_{-2.0}
\pm 1.6$ MeV. The $\Omega(2012)$ is the first $\Omega$ excited state
with preferred negative parity~\cite{Yelton:2018mag} and it is a PDG
three-star state. Before this new observation, there is only one
three-star $\Omega$ resonance, $\Omega(2250)$, with its spin-parity
unknown. Further investigations about the $\Omega$ excited states
are mostly welcome.

On the theoretical side, there exist several quark model studies of
the $\Omega$ excited states before the Belle observation. A pioneer
work was done in Ref.~\cite{Chao:1980em}, where an $\Omega$
resonance was predicted with mass about 2020 MeV and spin-parity
$J^P = 3/2^-$. The excitation spectrum for multistrange baryons were
investigated in Ref.~\cite{Pervin:2007wa} in a constituent quark
model, where a $3/2^-$ $\Omega$ excited state was obtained with mass
about $1953$ MeV. Within the extended chiral quark model, the
$\bar{K}\Xi$ and $\omega \Omega$ interactions were studied in
Refs.~\cite{Wang:2007bf,Wang:2008zzz}, in which, the $\Omega$
excited states in the $\bar{K}\Xi$ system with $J^P =1/2^-$ and
$\omega\Omega$ system with $J^P = 3/2^-$ or $5/2^-$, were obtained.
These $\Omega$ excited states with negative parity have also been
studied by using an extended quark model in the five quark
picture~\cite{Yuan:2012zs,An:2013zoa,An:2014lga}. It was found that
the lowest $3/2^-$ state has a mass around $1785 \pm 25$ MeV, which
is lower than the one of the lowest $1/2^-$ state~\cite{An:2014lga}.
This indicates that five-quark components are dominant in the wave
functions of those $\Omega$ resonances with lower
masses~\cite{An:2013zoa,An:2014lga}.

After its discovery the $\Omega(2012)$ resonance was studied in the
framework of QCD sum rules in
Refs.~\cite{Aliev:2018syi,Aliev:2018yjo}, where the $\Omega(2012)$
can be interpreted as a $1P$ orbital excitation of the ground
$\Omega$ baryon with $J^P = 3/2^-$. The two body strong decays of
$\Omega(2012)$ resonance were also studied in
Refs.~\cite{Xiao:2018pwe,Wang:2018hmi,Liu:2019wdr}, within a
non-relativistic constituent quark potential model, in which it was
found that the strong decay of $\Omega(2012)$ is predominated by
$\bar{K}\Xi$ mode.

Furthermore, the topic of hadronic molecular states, with mesons and
baryons bound by strong interactions in $s$-wave, has been well
developed by the combination of the chiral Lagrangians with
nonperturbative unitary techniques in coupled channels, which has
been a very fruitful scheme to study the nature of many baryon
resonances~\cite{Jido:2003cb,Magas:2005vu,Hyodo:2007jq,Xie:2017gwc}.
The analysis of meson-baryon scattering amplitudes shows poles,
which can be identified with existing baryon resonances or new ones.
In this way the $\Omega$ resonances are dynamically
generated~\cite{Kolomeitsev:2003kt,Sarkar:2004jh,GarciaRecio:2006bk,Si-Qi:2016gmh}
from the coupled channels interactions of the $\bar{K}\Xi^*(1530)$
and $\eta \Omega$ in $s$-wave.

Indeed, there is growing evidence that the newly observed
$\Omega(2012)$ can be interpreted as a hadronic molecular state,
with $J^P = 3/2^-$, as discussed in
Refs.~\cite{Valderrama:2018bmv,Lin:2018nqd,Pavao:2018xub,Huang:2018wth,Polyakov:2018mow,Lin:2019tex}.
However, the large decay width for $\Omega(2012)\to \bar{K}\Xi^* \to
\bar{K}\pi \Xi$, predicted by the molecular
nature~\cite{Valderrama:2018bmv,Lin:2018nqd,Pavao:2018xub}, is in
disagreement with the very recent Belle
measurement~\cite{Jia:2019eav}. The measured ratio of the three body
decay width to the one of the two body decay,
$R^{\bar{K}\pi\Xi}_{\bar{K}\Xi} = \Gamma_{\Omega(2012) \to
\bar{K}\pi \Xi}/ \Gamma_{\Omega(2012) \to {\bar K}\Xi}$, is less
than $11.9\%$ at the $90\%$ confidence level~\cite{Jia:2019eav}.
Based on this recent measurement,
Refs.~\cite{Lu:2020ste,Ikeno:2020vqv} claimed a reasonable
reproduction of the experimental data of the Belle
collaboration~\cite{Yelton:2018mag,Jia:2019eav}, and concluded that
the experimental data on the $\Omega(2012)$ are compatible with the
molecular picture and the theoretical results are rather stable with
different sets of model parameters of natural size.

The nonleptonic weak decays of charmed baryons can be useful tools
to study hadron
resonances~\cite{Oset:2016lyh,Hyodo:2011js,Miyahara:2015cja,Xie:2016evi,Xie:2017xwx,Xie:2017erh,Xie:2017mbe,Liu:2019dqc}.
The double strange baryon $\Xi^*(1620)^0$ was firstly observed in
its decay mode to $\pi^+\Xi^-$ via $\Xi^+_c \to \pi^+\pi^+\Xi^-$
process, measured by the Belle
collaboration~\cite{Sumihama:2018moz}. In Ref.~\cite{Lu:2016ogy} the
role of the $\Lambda^+_c \to \pi^+ n \bar{K}^0$ decay in testing
$SU(3)$ flavor symmetry and final state interactions was
investigated. Taking the advantage of these ideas and the previous
works of Refs.~\cite{Lu:2020ste,Ikeno:2020vqv}, we study the
$\Omega(2012)$ resonance~\footnote{It is worth to mention that, in
this work, the $\Omega(2012)$ is a dynamically generated state with
$J^P = 3/2^-$ from the coupled channels interactions of the
$\bar{K}\Xi^*(1530)$ and $\eta \Omega$ in $s$-wave and $\bar{K}\Xi$
in $d$-wave~\cite{Lu:2020ste,Ikeno:2020vqv}.} in the $\Omega^0_c \to
\pi^+ \bar{K}\Xi^*(1530)(\eta \Omega) \to \pi^+ (\bar{K}\Xi)^-$ and
$\pi^+ (\bar{K}\Xi\pi)^-$ decays, showing that they provide a good
filter for $I=0$ and strangeness $S=-3$ resonances, which can be
used to probe the nature of the $\Omega$ excited states.

The paper is organized as follows. In the next section, we present
the formalism and ingredients of the decay amplitudes of the three
and four-body decays of $\Omega^0_c$. Numerical results are given in
Section III, followed by a short summary in the last section.

\section{Formalism and ingredients}

\begin{figure}[htbp]
\centering
  \includegraphics[scale=0.6]{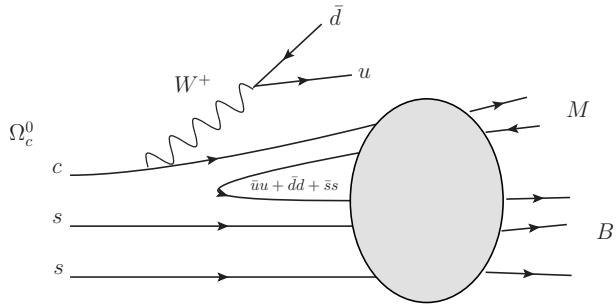}\\
  \caption{Dominant quark-line schematic diagram for $\Omega_c^0 \to \pi^+ {\rm MB}$ decay.}\label{Fig:quarkline}
\end{figure}

Following Refs.~\cite{Miyahara:2015cja,Xie:2016evi}, the Cabibbo
favored process of $\Omega^0_c$ into a $\pi^+$ plus a pair of ground
state pseudoscalar mesons and decuplet baryons (MB) is as follows:
in the first step the charmed quark in $\Omega^0_c$ turns into a
strange quark with a $\pi^+$($u\bar{d}$ pair) by the weak decay as
shown in Fig.~\ref{Fig:quarkline}. Then the $sss$ cluster hadronizes
with a new $\bar{q}q$ pair with the quantum numbers $J^{PC} =
0^{++}$ of the vacuum, to form a pair of meson and baryon. Finally,
the final-state interactions of the MB leads to the dynamical
generation of the $\Omega(2012)$.

\subsection{Decay amplitudes}

As in Refs.~\cite{Miyahara:2015cja,Xie:2016evi}, one can easily
obtain the final meson-baryon states as

\begin{align}\label{Eq:mbproduction}
|{\rm MB}\big>=&|s(\bar{u}u+\bar{d}d+\overline{s}s)ss\rangle
\notag\\=& \frac{1}{\sqrt{3}} \left(|K^-\Xi^{*0}\rangle +
|\bar{K}^0\Xi^{*-}\rangle \right) -
\frac{1}{\sqrt{3}}|\eta\Omega\rangle,
\notag\\=&\sqrt{\frac{2}{3}}|\bar{K}\Xi^*\rangle_{I=0} -
\frac{1}{\sqrt{3}}|\eta\Omega\rangle,
\end{align}
where the last step is obtained in the isospin basis using the
convention of Ref.~\cite{Oset:1997it}: $|K^-\rangle =
-|\frac{1}{2}-\frac{1}{2}\rangle$, and the flavor states of the
baryons and $\eta$ meson are as
follows~\cite{Pavao:2018xub,Miyahara:2015cja}:
\begin{eqnarray}
|\Xi^{*0}\rangle &=& \frac{1}{\sqrt{3}} |uss + sus + ssu \rangle, \\
|\Xi^{*-}\rangle &=&  \frac{1}{\sqrt{3}} |dss + sds + ssd \rangle, \\
|\Omega \rangle &=& |sss \rangle, \\
|\eta \rangle &=& \frac{1}{\sqrt{3}} |\bar{u}u + \bar{d}d - \bar{s}s
\rangle.
\end{eqnarray}

\begin{figure}[htbp]
\begin{center}
\includegraphics[scale=0.7]{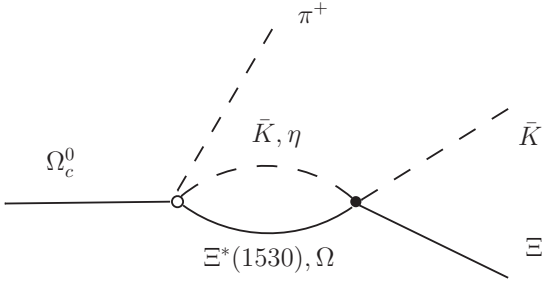}
\caption{Diagram for the meson-baryon final-state interaction for
the $\Omega^0_c \to \pi^+ \Omega(2012)^- \to \pi^+ (\bar{K} \Xi)^-$
decay.} \label{Fig:mbfsi}
\end{center}
\end{figure}

After the production of $\bar{K}\Xi^*(1530)$ and $\eta \Omega$
pairs, the final-state interactions between the mesons ($\bar{K}$,
$\eta$) and the baryons [$\Xi^*(1530)$, $\Omega$] take place, which
can be parameterized by the re-scattering shown in
Fig.~\ref{Fig:mbfsi} at the hadronic level for the production of
$\Omega(2012)$, and then it decays into $\bar{K}\Xi$ in $d$-wave.

\begin{table*}[htbp]
\centering \caption{Pole positions
$(M_{\Omega^*},\Gamma_{\Omega^*})$ of the $\Omega(2012)$ and the
couplings to different channels obtained with $q_{\rm max}$ (see
more details in Ref.~\cite{Lu:2020ste}).}\label{parameters}
\begin{tabular}{c|c|c|c|c|c|c}
\hline\hline Model & $\Lambda = q_{\rm max}$ (MeV) & $M_{\Omega^*}$ (${\rm MeV}$) & $\Gamma_{\Omega^*}$ (${\rm MeV}$)  & $g_{\Omega^*\bar{K}\Xi^*}$  & $g_{\Omega^*\eta\Omega}$ & $g_{\Omega^*\bar{K}\Xi}$   \\
\hline
I   & $735$   & $2012.3$   & $8.3$   & $(1.826, -0.064)$  & $(3.350,  0.159)$  & $(-0.419, -0.040)$ \\
II  & $750$   & $2012.2$   & $7.8$   & $(1.796, -0.128)$  & $(3.448,  0.298)$  & $(-0.399, -0.109)$ \\
III & $800$   & $2012.4$   & $6.4$   & $(1.574,  0.188)$  & $(3.590, -0.313)$  & $(-0.307,  0.201)$ \\
IV  & $850$   & $2012.4$   & $6.4$   & $(1.386,  0.090)$  & $(3.777, -0.151)$  & $(-0.353,  0.109)$ \\
V   & $900$   & $2012.4$   & $6.4$   & $(1.251,  0.063)$  & $(3.853, -0.111)$  & $(-0.363,  0.082)$ \\
\hline \hline
\end{tabular}
\end{table*}

According to Eq.~\eqref{Eq:mbproduction}, we can write down the
$\Omega^0_c \to \pi^+ \bar{K} \Xi$ decay amplitude of
Fig.~\ref{Fig:mbfsi} as,
\begin{align}\label{amtitu2}
\mathcal{M}_{\Omega_c^0\to\pi\bar{K}\Xi}=& V_p \left( \sqrt{\frac{2}{3}}G_{\bar{K}\Xi^*}(M_{\rm inv})t_{\bar{K}\Xi^*\to\bar{K}\Xi} (M_{\rm inv}) \notag \right .\\
& \left. -\sqrt{\frac{1}{3}}G_{\eta\Omega}(M_{\rm inv})
t_{\eta\Omega\to\bar{K}\Xi} (M_{\rm inv}) \right),
\end{align}
with $M_{\bar{K}\Xi}$ the invariant mass of $\bar{K}\Xi$. Similarly,
one can obtain the decay amplitude for $\Omega^0_c \to \pi^+ \bar{K}
\Xi^*(1530)$ as,
\begin{align}\label{amtitu1}
\mathcal{M}_{\Omega_c^0\to\pi\bar{K}\Xi^*}=&V_p \left ( \sqrt{\frac{2}{3}}[1+G_{\bar{K}\Xi^*} (M_{\rm inv}) t_{\bar{K}\Xi^*\to\bar{K}\Xi^*}(M_{\rm inv})]\notag \right .\\
& \left . -\sqrt{\frac{1}{3}}G_{\eta\Omega} (M_{\rm inv}
)t_{\eta\Omega\to\bar{K}\Xi^*} (M_{\rm inv}) \right ).
\end{align}
where the factor $V_P$ is assumed to be constant in the relevant
energy region~\cite{Miyahara:2015cja,Liang:2014tia,Xie:2018rqv}, and
its actual value should be determined from the experimental
measurements for a certain decay channel. The loop functions
$G_{\bar{K}\Xi^*}$ and $G_{\eta \Omega}$ depend on the invariant
mass, $M_{\rm inv}$, of the final $\bar{K} \Xi$ or
$\bar{K}\Xi^*(1530)$ system. The two body scattering amplitudes
$t_{\bar{K}\Xi^* \to \bar{K} \Xi(\Xi^*)}$ and $t_{\eta \Omega \to
\bar{K}\Xi(\Xi^*) }$ are those obtained in the chiral unitary
approach, which depend also on $M_{\rm inv}$, and we take them as,
\begin{align}
t_{\bar{K}\Xi^*\to\bar{K}\Xi^*}=&\frac{g_{\Omega^*\bar{K}\Xi^*}g_{\Omega^*\bar{K}\Xi^*}}{M_{\rm inv} -M_{\Omega^*}+i\Gamma_{\Omega^*}/2}, \\
t_{\eta\Omega\to\bar{K}\Xi^*}=&\frac{g_{\Omega^*\eta\Omega}g_{\Omega^*\bar{K}\Xi^*}}{M_{\rm inv} -M_{\Omega^*}+i\Gamma_{\Omega^*}/2}, \\
t_{\bar{K}\Xi^*\to\bar{K}\Xi}=&\frac{g_{\Omega^*\bar{K}\Xi^*}g_{\Omega^*\bar{K}\Xi}}{M_{\rm inv} -M_{\Omega^*}+i\Gamma_{\Omega^*}/2}, \\
t_{\eta\Omega\to\bar{K}\Xi}=&\frac{g_{\Omega^*\eta\Omega}g_{\Omega^*\bar{K}\Xi}}{M_{\rm
inv} -M_{\Omega^*}+i\Gamma_{\Omega^*}/2},
\end{align}
where these coupling constants, the mass and width of the
$\Omega(2012)$ are obtained in Ref.~\cite{Lu:2020ste} with different
sets of parameters as shown in Table~\ref{parameters}.

\begin{figure}[htbp]
\centering
\includegraphics[scale=0.6]{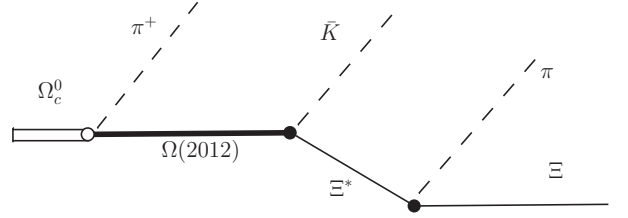}\\
\caption{Schematic diagram for the decay of $\Omega^0_c \to \pi^+
\Omega(2012)^- \to \pi^+ [\bar{K}\Xi^*(1530)]^- \to \pi^+ (\bar{K}
\Xi \pi)^-$. }\label{Fig:4body}
\end{figure}

Next we consider the $\Omega^0_c \to \pi^+ \Omega(2012)^- \to \pi^+
[\bar{K}\Xi^*(1530)]^- \to \pi^+ (\bar{K} \Xi \pi)^-$ decay, as
shown in Fig.~\ref{Fig:4body}, where the $\Omega(2012)$ is produced
by the final state interactions of $\bar{K}\Xi^*(1530)$ and $\eta
\Omega$ in coupled channels. The decay amplitude of $\Omega^0_c \to
\pi^+ \bar{K}\Xi\pi$ decay can be written as
\begin{align}
\mathcal{M}_{\Omega_c^0\to\pi^+\bar{K}\Xi\pi} =
\frac{g_{\Xi^*\Xi\pi}\bar{p}_{\pi}
\mathcal{M}_{\Omega_c^0\to\pi^+\bar{K}\Xi^*}
}{M_{\Xi\pi}-M_{\Xi^*}+i\Gamma_{\Xi^*}/2},
\end{align}
where $M_{\Xi \pi}$ is the invariant mass of $\Xi \pi$ system, and
$\Gamma_{\Xi^*}$ is energy dependent, and its explicit form is given
by~\footnote{The $p$-wave $\Xi^*(1530) \to \Xi \pi$ decay is
considered.}
\begin{align}
\Gamma_{\Xi^*}=\frac{1}{2\pi}\frac{M_{\Xi}}{M_{\Xi
\pi}}&\bar{p}_{\pi}(g_{\Xi^*\Xi\pi}\bar{p}_{\pi})^2=\Gamma^{\rm
on}_{\Xi^*}\frac{M_{\Xi^*}}{M_{\Xi\pi}}(\frac{\bar{p}_\pi}{\bar{p}_\pi^{~\rm
on}})^3,  \label{eq:gamaXistar}
\end{align}
where
\begin{align}
\bar{p}_{\pi}=&\frac{\sqrt{[M^2_{\Xi\pi}-(M_{\Xi} + m_{\pi})^2][M^2_{\Xi\pi}-(M_{\Xi} - m_{\pi})^2]}}{2M_{\Xi\pi}},\notag\\
\bar{p}^{~\rm on}_{\pi}=&\frac{\sqrt{[M^2_{\Xi^*}-(M_{\Xi} +
m_{\pi})^2][M^2_{\Xi^*}-(M_{\Xi} - m_{\pi})^2]}}{2M_{\Xi^*}}.\notag
\end{align}

On the other hand, the coupling constant $g_{\Xi^*\Xi \pi} = 4.4
\times 10^{-3}$ ${\rm MeV}^{-1}$ can be easily obtained from
Eq.~\eqref{eq:gamaXistar} with the values of $m_\pi = 138.04$,
$M_{\Xi^*} = 1533.4$, $M_{\Xi} = 1318.29$ MeV, and $\Gamma^{\rm
on}_{\Xi^*} = 9.5$ MeV.

\subsection{Invariant mass distributions}

With all the ingredients obtained above, one can write down the
invariant mass distributions for the $\Omega^0_c \to \pi^+
\Omega(2012)^- \to \pi^+ \bar{K}\Xi$ decay
as~\cite{Tanabashi:2018oca}
\begin{align}
\frac{d\Gamma_{\Omega^0_c\to \pi^+ \bar{K}\Xi}}{dM_{\bar{K} \Xi}} =
&\frac{1}{16\pi^3}\frac{M_{\Xi}}{M_{\Omega^0_c}}
p^3_{\pi}p_{\bar{K}} \sum|{\mathcal{M}}_{\Omega^0_c \to \pi^+
\bar{K}\Xi}|^2, \label{eq:dGammadMKbarXi}
\end{align}
where we have only considered $L=1$ for the $\pi^+$ in the
$\Omega^0_c(1/2^+) \to \pi^+(0^-) \Omega(2012)^-(3/2^-)$ transition
to match angular momentum conservation~\footnote{Since the
transition of $\Omega^0_c(1/2^+) \to \pi^+(0^-)
\Omega(2012)^-(3/2^-)$ is weak decay, in general, the $d$-wave term
also gives contribution, which is neglected in this work, because it
is suppressed at the low energy region.}, and
\begin{align}\label{ppi}
p_{\pi} = \frac{\sqrt{[M_{\Omega_c^0}^2-(m_{\pi}+ M_{\bar{K} \Xi}
)^2][M_{\Omega_c^0}^2-(m_{\pi}-M_{\bar{K}
\Xi})^2]}}{2M_{\Omega_c^0}}.\notag
\end{align}
\begin{align}
p_{\bar{K}}=\frac{\sqrt{[M^2_{\bar{K}
\Xi}-(m_{\bar{K}}+M_{\Xi})^2][M^2_{\bar{K}
\Xi}-(m_{\bar{K}}-M_{\Xi})^2]}}{2M_{\Omega_c^0}}\notag.
\end{align}

Similarly, the $d\Gamma_{\Omega^0_c \to \pi^+
\bar{K}\Xi^*}/dM_{\bar{K} \Xi^*}$ can be easily obtained by applying
the substitution to $d\Gamma_{\Omega^0_c\to \pi^+
\bar{K}\Xi}/dM_{\bar{K} \Xi}$ with $M_{\Xi} \to M_{\Xi^*}$,
$M_{\bar{K}\Xi} \to M_{\bar{K}\Xi^*}$, and ${\cal M}_{\Omega^0_c \to
\pi^+ \bar{K}\Xi} \to {\cal M}_{\Omega^0_c \to \pi^+ \bar{K}\Xi^*}$.

For the $\Omega^0_c \to \pi^+ \Omega(2012)^- \to \pi^+ \bar{K}
\Xi\pi$ decay, the $\bar{K}\Xi \pi$ invariant mass distribution is
given by~\cite{Xie:2018gbi,Jing:2020tth},

\begin{align}
\frac{d\Gamma_{\Omega^0_c\to
\pi\bar{K}\Xi\pi}}{dM_{\bar{K}\Xi\pi}dM_{\Xi\pi}} = & \frac{M_{\Xi}
p'_{\pi} \tilde{p}_K \bar{p}_\pi}{64\pi^5 M_{\Omega_c^0}}  \sum
|{\mathcal{M}}_{\Omega_c^0\to\pi^+ \bar{K}\Xi \pi}|^2,
\end{align}
with
\begin{align}
p'_{\pi} = \frac{\sqrt{[M_{\Omega_c^0}^2-(m_{\pi}+ M_{\bar{K}
\Xi\pi} )^2][M_{\Omega_c^0}^2-(m_{\pi}-M_{\bar{K} \Xi
\pi})^2]}}{2M_{\Omega_c^0}}, \notag
\end{align}
\begin{align}
{\tilde
p}_{\bar{K}}=\frac{\sqrt{[M^2_{\bar{K}\Xi\pi}-(m_{\bar{K}}+M_{\Xi\pi})^2][M^2_{\bar{K}\Xi\pi}
- (m_{\bar{K}}-M_{\Xi\pi})^2]}}{2M_{\bar{K}\Xi\pi}}. \notag
\end{align}

After the integration of $M_{\pi \Xi}$, we can obtain
\begin{align}
\frac{d\Gamma_{\Omega^0_c\to \pi\bar{K}\Xi\pi}}{dM_{\bar{K}\Xi \pi}}
= &\int^{M_{\bar{K}\Xi \pi} - m_{\bar{K}}}_{M_{\Xi} +
m_{\pi}}\frac{d\Gamma_{\Omega^0_c\to
\pi\bar{K}\Xi\pi}}{dM_{\Xi\pi}dM_{\bar{K}\Xi \pi}}dM_{\Xi\pi} .
\end{align}

\section{Numerical results}

In this section, we show our theoretical predictions for the
production of $\Omega(2012)$ in different final states in the
$\Omega^0_c$ decay. Note that the physical masses of the involved
particles are taken from PDG \cite{Tanabashi:2018oca} and we take
the isospin averaged values for $m_{K} = 495.64$, $m_\eta = 547.86$,
$m_{\Omega} = 1672.45$, and $M_{\Omega^0_c} = 2695.2$ MeV. In
addition, the following numerical results are obtained with $V_P =
1$.

\begin{figure}[htbp]
\centering
  \includegraphics[scale=0.33]{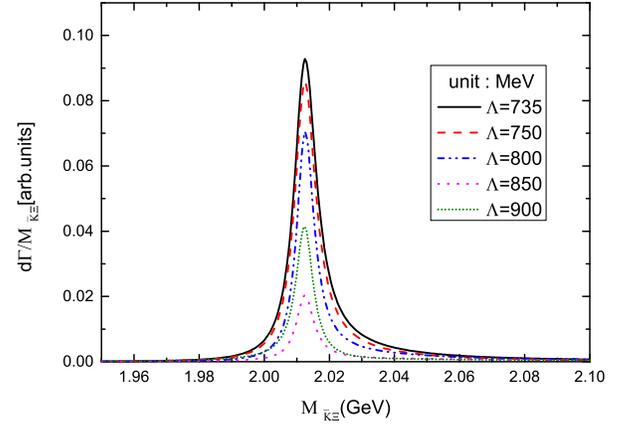}\\
  \caption{$\bar{K}\Xi$ invariant mass distributions of the $\Omega^0_c \to \pi^+ (\bar{K}\Xi)^-$ decay . }\label{Fig:all3body}
\end{figure}

\begin{figure}[htbp]
\centering
  \includegraphics[scale=0.33]{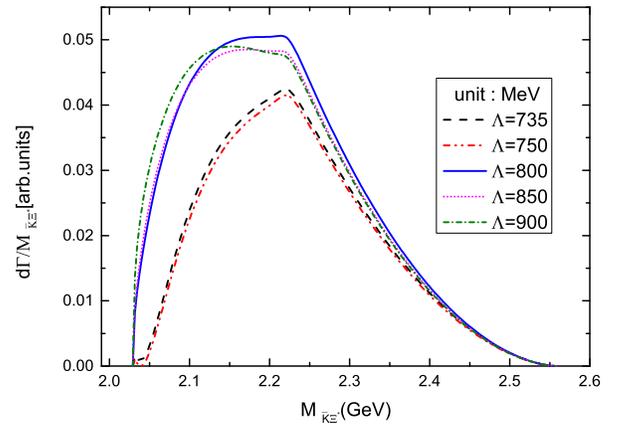}\\
  \caption{$\bar{K}\Xi^*$ invariant mass distribution of the $\Omega^0_c \to \pi^+ (\bar{K} \Xi^*)^-$ decay. }\label{Fig:all3bodyst}
\end{figure}

We first show in Figs.~\ref{Fig:all3body} and \ref{Fig:all3bodyst}
the theoretical predictions for the invariant mass distribution
$d\Gamma/dM_{\bar{K}\Xi}$ and $d\Gamma/dM_{\bar{K}\Xi^*(1530)}$.
From Fig.~\ref{Fig:all3body} one can see clearly the shape of the
$\Omega(2012)$, while from the invariant mass distribution of
$\bar{K}\Xi^*(1530)$ as shown in Fig.~\ref{Fig:all3bodyst}, there is
no signal of the $\Omega(2012)$ resonance, this is because its mass
is below the $\bar{K}\Xi^*$ mass threshold, and its width is narrow.
The interference between the tree level contribution and the final
state interactions of $\bar{K}\Xi^*(1530)$ and $\eta \Omega$ makes
the production of $\Omega(2012)$ in the $\bar{K}\Xi^*(1530)$ channel
even worse. In other words, the first term in Eq.~\eqref{amtitu1}
contributes at the tree level to the $\Omega^0_c \to \pi^+ \bar{K}
\Xi^*(1530)$ decay, but does not contribute to the production of
$\Omega(2012)$, followed by decaying into $\bar{K}\Xi^*(1530)$.

\begin{figure}[htbp]
\centering
  \includegraphics[scale=0.33]{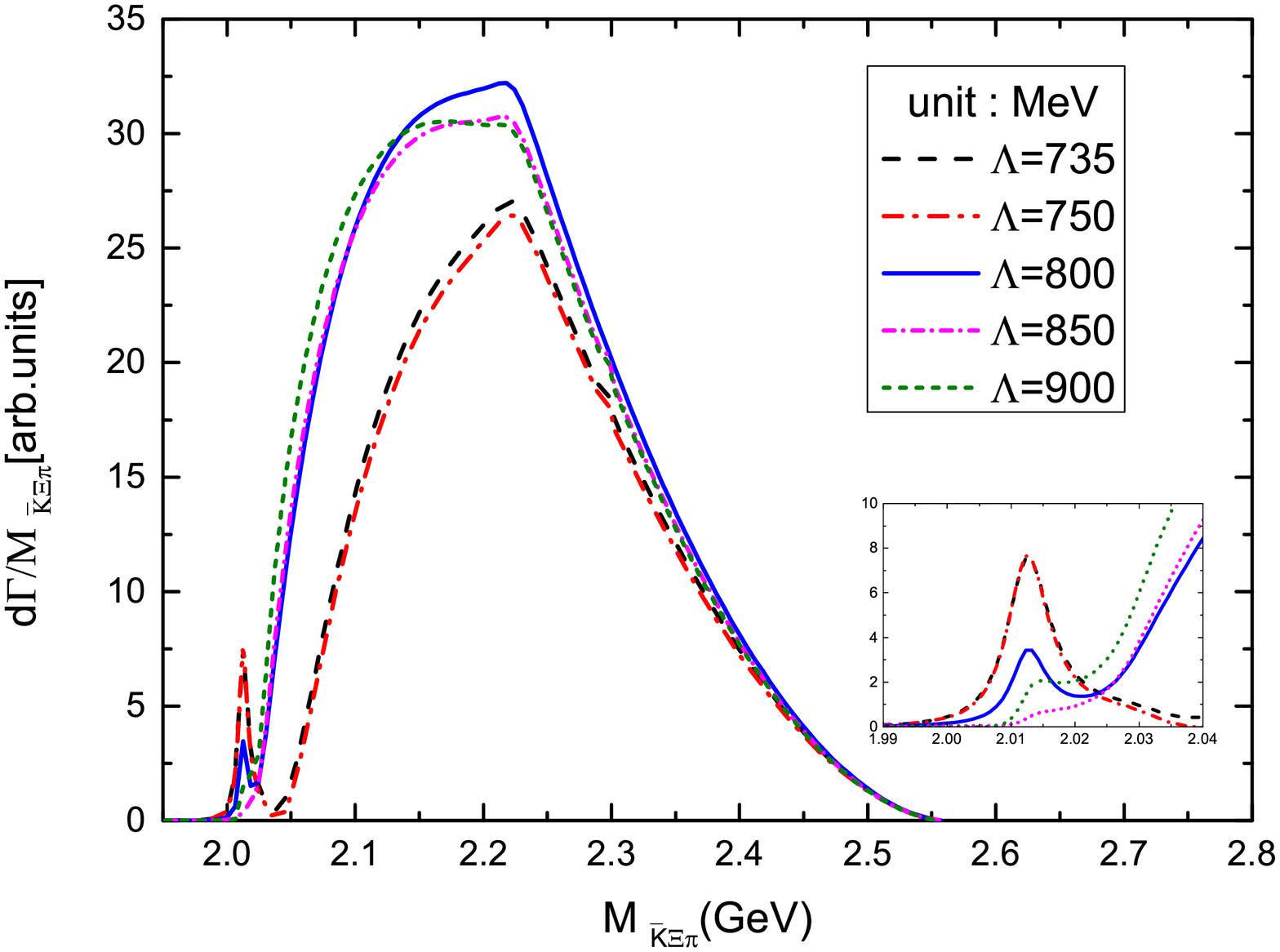}\\
  \caption{$\bar{K}\Xi \pi$ invariant mass distributions of the $\Omega^0_c \to\pi^+ \bar{K}\Xi^*(1530) \to \pi^+ \bar{K}\Xi\pi$ decay . }\label{Fig:all4body}
\end{figure}

Next we consider the $\Omega^0_c \to \pi^+ \bar{K}\Xi^*(1530) \to
\pi^+ \bar{K}\Xi \pi$ decay. The invariant mass distribution
$d\Gamma/dM_{\bar{K}\Xi\pi}$ is shown in Fig.~\ref{Fig:all4body},
where one can see that the tree level contributions provide a very
big background, and, hence, the signal of the $\Omega(2012)$ is
rather weak. However, the $\Omega(2012)$ may give significant
contribution to the invariant mass distributions of $\bar{K}\Xi \pi$
close to threshold, as shown in the sub-figure of
Fig.~\ref{Fig:all4body}, especially for the Models I, II and III.

In general we cannot fix the value of $V_P$, which should be
determined by experimental measurements. Therefore, it would be
interesting to remove the uncertainties arising from the
introduction of $V_P$ by investigating the ratios between different
partial decay widths, where the effect of the $V_P$ factor is
canceled, and which only reflect the production of $\Omega(2012)$
through the final state interactions of $\bar{K}\Xi^*(1530)$ and
$\eta \Omega$. For such a purpose, we define

\begin{align}
R^{\bar{K}\Xi\pi}_{\bar{K}\Xi} = &\frac{\Gamma[\Omega^0_c\to \pi^+\Omega(2012)^- \to\pi^+\bar{K}\Xi\pi]}{\Gamma[\Omega^0_c \to \pi^+ \Omega(2012)^- \to\pi^+\bar{K}\Xi]} \notag\\
=&\frac{\int^{M_{\Omega^*} + 2\Gamma_{\Omega^*}}_{M_{\Omega^*} -
2\Gamma_{\Omega^*}}\frac{d\Gamma_{\Omega^0_c \to \pi^+
\bar{K}\Xi\pi}}{dM_{\bar{K}\Xi\pi}}dM_{\bar{K}\Xi\pi}}{\int^{M_{\Omega^*}
+ 2\Gamma_{\Omega^*}}_{M_{\Omega^*} -
2\Gamma_{\Omega^*}}\frac{d\Gamma_{\Omega^0_c \to
\pi^+\bar{K}\Xi}}{dM_{\bar{K}\Xi}}dM_{\bar{K}\Xi}}. \label{eq:ratio}
\end{align}

As discussed before, the main contribution to the $\Omega^0_c \to
\pi^+ \bar{K}\Xi\pi$ decay comes from the tree level of the
production of $\bar{K}\Xi^*(1530)$. In order to consider the
contribution from the $\Omega(2012)$ resonance, we need to switch
off the first term in Eq.~\eqref{amtitu1}, which means that we have
to change the ${\cal M}_{\Omega_c^0\to\pi\bar{K}\Xi^*}$ in the
following form
\begin{align}\label{amtitu3}
& {\cal M}_{\Omega_c^0\to\pi\bar{K}\Xi^*} \to  \tilde{{\cal M}}_{\Omega_c^0\to\pi\bar{K}\Xi^*} =\notag\\
& V_p \left (
\sqrt{\frac{2}{3}}G_{\bar{K}\Xi^*}t_{\bar{K}\Xi^*\to\bar{K}\Xi^*}-\sqrt{\frac{1}{3}}G_{\eta\Omega}t_{\eta\Omega\to\bar{K}\Xi^*}
\right ).
\end{align}

Furthermore, as shown in Eq.~\eqref{eq:ratio}, to get the partial
decay widths of $\Gamma[\Omega^0_c\to \pi^+\Omega(2012)^-
\to\pi^+\bar{K}\Xi\pi]$ and $\Gamma[\Omega^0_c\to
\pi^+\Omega(2012)^- \to\pi^+\bar{K}\Xi\pi]$, we have integrated
$dM_{\bar{K}\Xi\pi}$ and $dM_{\bar{K}\Xi}$ over the range of
$[M_{\Omega^*}-2\Gamma_{\Omega^*},M_{\Omega^*}+2\Gamma_{\Omega^*}]$,
in which the contributions from $\Omega(2012)$ resonance are
covered.

The numerical results for the ratio $R^{\bar{K}\Xi\pi}_{\bar{K}\Xi}$
are listed in Table~\ref{result1}. On can see that these predictions
are in agreement with the measurements reported by the Belle
collaboration~\cite{Jia:2019eav}, as expected. This ratio is
relevant because it is obtained with no free parameters (all the
model parameters are fixed by previous works) and, thus, it is a
prediction of the model. We expect that these numerical results
could be tested by future experimental measurements.

\begin{table}[htbp]
\centering \caption{Predicted ratio $R^{\bar{K}\Xi\pi}_{\bar{K}\Xi}$
for different cutoffs.}\label{result1}
\begin{tabular}{c|ccccc}
\hline
\hline
$\Lambda = q_{\rm max}$(MeV)& 735 & 750 & 800 & 850 & 900\\
\hline
$R^{\bar{K}\Xi\pi}_{\bar{K}\Xi} (\%) $ & 13.9 & 13.8 & 13.5 & 10.0 & 7.3 \\
\hline \hline
\end{tabular}
\end{table}

\section{Summary}

In summary, we have investigated the nonleptonic weak decays of
$\Omega_c^0 \to \pi^+ \bar{K}\Xi^*(1530) (\eta \Omega) \to \pi^+
(\bar{K}\Xi)^-$ as a tool to study the newly observed $\Omega(2012)$
resonance via the final-state interactions of the
$\bar{K}\Xi^*(1530)$ and $\eta \Omega$ pairs. We assume that the
weak interaction part is dominated by the Cabibbo favored charm
quark decay process: $c(ss) \to (s + u + \bar{d})(ss)$, then the
$sss$ cluster and a quark-antiquark pair from the vacuum hadronize
into the intermediate meson-baryon states, in this case, the
$\bar{K}\Xi^*(1530)$ and $\eta \Omega$. Accordingly, the final
$\bar{K}\Xi^*(1530)$ and $\eta \Omega$ states are in pure isospin $I
= 0$ combinations, and the final state interaction of
$\bar{K}\Xi^*(1530)$ and $\eta \Omega$ can produce the
$\Omega(2012)$ state by using the chiral unitary approach. After the
$\Omega(2012)$ is dynamically generated in the above process, it
will decay into $\bar{K}\Xi$ and $\bar{K}\Xi\pi$, and shows a peak
or bump structure in the $\bar{K}\Xi$ and $\bar{K}\Xi\pi$ invariant
mass distributions. Thus, we have calculated the invariant mass
distributions of $d\Gamma_{\Omega^0_c \to \pi^+
\bar{K}\Xi}/dM_{\bar{K}\Xi}$ and $d\Gamma_{\Omega^0_c \to \pi^+
\bar{K}\Xi\pi}/dM_{\bar{K}\Xi\pi}$. We have seen that the
$\Omega^0_c \to \pi^+ \bar{K}\Xi\pi$ decay is not well suited to
study the $\Omega(2012)$ resonance because the dominant contribution
is from the $\Omega^0_c \to \pi^+ \bar{K}\Xi^*(1530)$ decay at tree
level, which will not contribute to the production of
$\Omega(2012)$. However, the $\Omega(2012)$ peak can be clearly seen
in the $\bar{K}\Xi$ invariant mass distribution of the $\Omega^0_c
\to \pi^+ \bar{K}\Xi$ decay.

We have also calculated the ratio, $R^{\bar{K}\Xi\pi}_{\bar{K}\Xi} =
{\rm Br}[\Omega_c^0 \to \pi^+ \Omega(2012)^- \to \pi^+ (\bar{K}\Xi
\pi)^-] / {\rm Br}[\Omega_c^0 \to \pi^+ \Omega(2012)^- \to \pi^+
(\bar{K}\Xi)^-$]. The numerical results are in agreement with the
Belle measurements~\cite{Jia:2019eav}. The good agreement with
experimental data of the chiral unitary approach as shown by us
here, provides extra support to the picture of the $\Omega(2012)$ as
a dynamically generated resonance.

Finally, we would like to stress that the predictions here are very
qualitative, since the contributions from other resonances are
neglected. We hope that the theoretical calculations presented in
this work may stimulate experimental interest in exploring the
$\Omega(2012)$ resonance or other $\Omega$ excited states through
the $\Omega^0_c$ decays. The proposed mechanism can provide valuable
information on the nature of the $\Omega(2012)$ and can in principle
be tested by future experiments.

\section*{Acknowledgments}

This work is partly supported by the National Natural Science
Foundation of China under Grants Nos. 11735003, 11975041,
11961141004, and 11961141012, and the fundamental Research Funds for
the Central Universities. It is also supported by the Key Research
Projects of Henan Higher Education Institutions under No. 20A140027,
the Academic Improvement Project of Zhengzhou University. and the
Youth Innovation Promotion Association CAS (2016367).


\begin{thebibliography}{99}


  %\cite{Tanabashi:2018oca}
\bibitem{Tanabashi:2018oca}
  M.~Tanabashi {\it et al.} [Particle Data Group],
  %``Review of Particle Physics,''
  Phys.\ Rev.\ D {\bf 98}, 030001 (2018).
%  doi:10.1103/PhysRevD.98.030001
  %%CITATION = doi:10.1103/PhysRevD.98.030001;%%
  %3986 citations counted in INSPIRE as of 03 Mar 2020

%\cite{Yelton:2018mag}
\bibitem{Yelton:2018mag}
  J.~Yelton {\it et al.} [Belle Collaboration],
  %``Observation of an Excited $\Omega^-$ Baryon,''
  Phys.\ Rev.\ Lett.\  {\bf 121}, 052003 (2018).
%  doi:10.1103/PhysRevLett.121.052003
%  [arXiv:1805.09384 [hep-ex]].
  %%CITATION = doi:10.1103/PhysRevLett.121.052003;%%
  %21 citations counted in INSPIRE as of 19 Feb 2020

  %\cite{Chao:1980em}
\bibitem{Chao:1980em}
  K.~T.~Chao, N.~Isgur and G.~Karl,
  %``Strangeness -2 and -3 Baryons in a Quark Model With Chromodynamics,''
  Phys.\ Rev.\ D {\bf 23}, 155 (1981).
%  doi:10.1103/PhysRevD.23.155
  %%CITATION = doi:10.1103/PhysRevD.23.155;%%
  %126 citations counted in INSPIRE as of 06 Mar 2020

%\cite{Pervin:2007wa}
\bibitem{Pervin:2007wa}
  M.~Pervin and W.~Roberts,
  %``Strangeness -2 and -3 baryons in a constituent quark model,''
  Phys.\ Rev.\ C {\bf 77}, 025202 (2008).
%  doi:10.1103/PhysRevC.77.025202
%  [arXiv:0709.4000 [nucl-th]].
  %%CITATION = doi:10.1103/PhysRevC.77.025202;%%
  %36 citations counted in INSPIRE as of 01 Jul 2020

%\cite{Wang:2007bf}
\bibitem{Wang:2007bf}
  W.~L.~Wang, F.~Huang, Z.~Y.~Zhang, Y.~W.~Yu and F.~Liu,
  %``Omega $\omega$ states in a chiral quark model,''
  Commun.\ Theor.\ Phys.\  {\bf 48}, 695 (2007).
%  doi:10.1088/0253-6102/48/4/025
  %%CITATION = doi:10.1088/0253-6102/48/4/025;%%
  %6 citations counted in INSPIRE as of 18 Jun 2020

%\cite{Wang:2008zzz}
\bibitem{Wang:2008zzz}
  W.~L.~Wang, F.~Huang, Z.~Y.~Zhang and F.~Liu,
  %``Xi anti-K interaction in a chiral model,''
  J.\ Phys.\ G {\bf 35}, 085003 (2008).
%  doi:10.1088/0954-3899/35/8/085003
  %%CITATION = doi:10.1088/0954-3899/35/8/085003;%%
  %13 citations counted in INSPIRE as of 18 Jun 2020

%\cite{Yuan:2012zs}
\bibitem{Yuan:2012zs}
  S.~G.~Yuan, C.~S.~An, K.~W.~Wei, B.~S.~Zou and H.~S.~Xu,
  %``Spectrum of low-lying $s^{3}Q\bar{Q}$ configurations with negative parity,''
  Phys.\ Rev.\ C {\bf 87}, 025205 (2013).
%  doi:10.1103/PhysRevC.87.025205
%  [arXiv:1208.1742 [hep-ph]].
  %%CITATION = doi:10.1103/PhysRevC.87.025205;%%
  %14 citations counted in INSPIRE as of 02 Mar 2020

%\cite{An:2013zoa}
\bibitem{An:2013zoa}
  C.~S.~An, B.~C.~Metsch and B.~S.~Zou,
  %``Mixing of the low-lying three- and five-quark $\Omega$ states with negative parity,''
  Phys.\ Rev.\ C {\bf 87}, 065207 (2013).
%  doi:10.1103/PhysRevC.87.065207
%  [arXiv:1304.6046 [hep-ph]].
  %%CITATION = doi:10.1103/PhysRevC.87.065207;%%
  %14 citations counted in INSPIRE as of 02 Mar 2020

%\cite{An:2014lga}
\bibitem{An:2014lga}
  C.~S.~An and B.~S.~Zou,
  %``Low-lying $\Omega$ states with negative parity in an extended quark model with Nambu-Jona-Lasinio interaction,''
  Phys.\ Rev.\ C {\bf 89}, 055209 (2014).
%  doi:10.1103/PhysRevC.89.055209
%  [arXiv:1403.7897 [hep-ph]].
  %%CITATION = doi:10.1103/PhysRevC.89.055209;%%
  %12 citations counted in INSPIRE as of 02 Mar 2020

%\cite{Aliev:2018syi}
\bibitem{Aliev:2018syi}
  T.~M.~Aliev, K.~Azizi, Y.~Sarac and H.~Sundu,
  %``Interpretation of the newly discovered $\Omega$(2012),''
  Phys.\ Rev.\ D {\bf 98}, 014031 (2018).
%  doi:10.1103/PhysRevD.98.014031
%  [arXiv:1806.01626 [hep-ph]].
  %%CITATION = doi:10.1103/PhysRevD.98.014031;%%
  %14 citations counted in INSPIRE as of 01 Jul 2020

%\cite{Aliev:2018yjo}
\bibitem{Aliev:2018yjo}
  T.~M.~Aliev, K.~Azizi, Y.~Sarac and H.~Sundu,
  %``Nature of the $\Omega (2012)$ through its strong decays,''
  Eur.\ Phys.\ J.\ C {\bf 78}, 894 (2018).
%  doi:10.1140/epjc/s10052-018-6375-y
%  [arXiv:1807.02145 [hep-ph]].
  %%CITATION = doi:10.1140/epjc/s10052-018-6375-y;%%
  %11 citations counted in INSPIRE as of 01 Jul 2020

 %\cite{Xiao:2018pwe}
\bibitem{Xiao:2018pwe}
  L.~Y.~Xiao and X.~H.~Zhong,
  %``Possible interpretation of the newly observed $\Omega$(2012) state,''
  Phys.\ Rev.\ D {\bf 98}, 034004 (2018).
%  doi:10.1103/PhysRevD.98.034004
%  [arXiv:1805.11285 [hep-ph]].
  %%CITATION = doi:10.1103/PhysRevD.98.034004;%%
  %13 citations counted in INSPIRE as of 02 Mar 2020


%\cite{Wang:2018hmi}
\bibitem{Wang:2018hmi}
  Z.~Y.~Wang, L.~C.~Gui, Q.~F.~L\"u, L.~Y.~Xiao and X.~H.~Zhong,
  %``Newly observed $\Omega(2012)$ state and strong decays of the low-lying $\Omega$ excitations,''
  Phys.\ Rev.\ D {\bf 98}, 114023 (2018).
%  doi:10.1103/PhysRevD.98.114023
%  [arXiv:1810.08318 [hep-ph]].
  %%CITATION = doi:10.1103/PhysRevD.98.114023;%%
  %4 citations counted in INSPIRE as of 02 Mar 2020

%\cite{Liu:2019wdr}
\bibitem{Liu:2019wdr}
  M.~S.~Liu, K.~L.~Wang, Q.~F.~L\"u and X.~H.~Zhong,
  %``$\Omega$ baryon spectrum and their decays in a constituent quark model,''
  Phys.\ Rev.\ D {\bf 101}, 016002 (2020).
%  doi:10.1103/PhysRevD.101.016002
%  [arXiv:1910.10322 [hep-ph]].
  %%CITATION = doi:10.1103/PhysRevD.101.016002;%%
  %3 citations counted in INSPIRE as of 02 Mar 2020

%\cite{Jido:2003cb}
\bibitem{Jido:2003cb}
  D.~Jido, J.~A.~Oller, E.~Oset, A.~Ramos and U.~G.~Mei\ss ner,
  %``Chiral dynamics of the two Lambda(1405) states,''
  Nucl.\ Phys.\ A {\bf 725}, 181 (2003).
%  doi:10.1016/S0375-9474(03)01598-7
%  [nucl-th/0303062].
  %%CITATION = doi:10.1016/S0375-9474(03)01598-7;%%
  %623 citations counted in INSPIRE as of 28 Jun 2020

%\cite{Magas:2005vu}
\bibitem{Magas:2005vu}
  V.~K.~Magas, E.~Oset and A.~Ramos,
  %``Evidence for the two pole structure of the Lambda(1405) resonance,''
  Phys.\ Rev.\ Lett.\  {\bf 95}, 052301 (2005).
%  doi:10.1103/PhysRevLett.95.052301
%  [hep-ph/0503043].
  %%CITATION = doi:10.1103/PhysRevLett.95.052301;%%
  %189 citations counted in INSPIRE as of 28 Jun 2020

%\cite{Hyodo:2007jq}
\bibitem{Hyodo:2007jq}
  T.~Hyodo and W.~Weise,
  %``Effective anti-K N interaction based on chiral SU(3) dynamics,''
  Phys.\ Rev.\ C {\bf 77}, 035204 (2008).
%  doi:10.1103/PhysRevC.77.035204
%  [arXiv:0712.1613 [nucl-th]].
  %%CITATION = doi:10.1103/PhysRevC.77.035204;%%
  %242 citations counted in INSPIRE as of 28 Jun 2020

%\cite{Xie:2017gwc}
\bibitem{Xie:2017gwc}
  J.~J.~Xie, W.~H.~Liang and E.~Oset,
  %``Hidden charm pentaquark and $\Lambda(1405)$ in the $\Lambda^0_b \to \eta_c K^- p (\pi \Sigma)$ reaction,''
  Phys.\ Lett.\ B {\bf 777}, 447 (2018).
%  doi:10.1016/j.physletb.2017.12.064
%  [arXiv:1711.01710 [hep-ph]].
  %%CITATION = doi:10.1016/j.physletb.2017.12.064;%%
  %6 citations counted in INSPIRE as of 28 Jun 2020

%\cite{Kolomeitsev:2003kt}
\bibitem{Kolomeitsev:2003kt}
  E.~E.~Kolomeitsev and M.~F.~M.~Lutz,
  %``On baryon resonances and chiral symmetry,''
  Phys.\ Lett.\ B {\bf 585}, 243 (2004).
%  doi:10.1016/j.physletb.2004.01.066
%  [nucl-th/0305101].
  %%CITATION = doi:10.1016/j.physletb.2004.01.066;%%
  %198 citations counted in INSPIRE as of 02 Mar 2020

%\cite{Sarkar:2004jh}
\bibitem{Sarkar:2004jh}
  S.~Sarkar, E.~Oset and M.~J.~Vicente Vacas,
  %``Baryonic resonances from baryon decuplet-meson octet interaction,''
  Nucl.\ Phys.\ A {\bf 750}, 294 (2005)
  Erratum: [Nucl.\ Phys.\ A {\bf 780}, 90 (2006)].
%  doi:10.1016/j.nuclphysa.2005.01.006, 10.1016/j.nuclphysa.2006.09.019
%  [nucl-th/0407025].
  %%CITATION = doi:10.1016/j.nuclphysa.2005.01.006, 10.1016/j.nuclphysa.2006.09.019;%%
  %159 citations counted in INSPIRE as of 02 Mar 2020


%\cite{GarciaRecio:2006bk}
\bibitem{GarciaRecio:2006bk}
  C.~Garc\'{i}a-Recio, J.~Nieves and L.~L.~Salcedo,
  %``Meson-Baryon s-wave Resonances with Strangeness -3,''
  Eur.\ Phys.\ J.\ A {\bf 31}, 540 (2007).
%  doi:10.1140/epja/i2006-10241-3
%  [hep-ph/0610353].
  %%CITATION = doi:10.1140/epja/i2006-10241-3;%%
  %6 citations counted in INSPIRE as of 13 Mar 2020

%\cite{Si-Qi:2016gmh}
\bibitem{Si-Qi:2016gmh}
  S.~Q.~Xu, J.~J.~Xie, X.~R.~Chen and D.~J.~Jia,
  %``The $\Xi^* \bar{K}$ and $\Omega \eta$ interaction within a chiral unitary approach,''
  Commun.\ Theor.\ Phys.\  {\bf 65}, 53 (2016).
%  doi:10.1088/0253-6102/65/1/53
%  [arXiv:1510.07419 [nucl-th]].
  %%CITATION = doi:10.1088/0253-6102/65/1/53;%%
  %5 citations counted in INSPIRE as of 02 Mar 2020

  %\cite{Pavao:2018xub}
\bibitem{Pavao:2018xub}
  R.~Pavao and E.~Oset,
  %``Coupled channels dynamics in the generation of the $\Omega (2012)$ resonance,''
  Eur.\ Phys.\ J.\ C {\bf 78}, 857 (2018).
%  doi:10.1140/epjc/s10052-018-6329-4
%  [arXiv:1808.01950 [hep-ph]].
  %%CITATION = doi:10.1140/epjc/s10052-018-6329-4;%%
  %6 citations counted in INSPIRE as of 19 Feb 2020


%\cite{Lin:2018nqd}
\bibitem{Lin:2018nqd}
  Y.~H.~Lin and B.~S.~Zou,
  %``Hadronic molecular assignment for the newly observed $\Omega^*$ state,''
  Phys.\ Rev.\ D {\bf 98}, 056013 (2018).
%  doi:10.1103/PhysRevD.98.056013
%  [arXiv:1807.00997 [hep-ph]].
  %%CITATION = doi:10.1103/PhysRevD.98.056013;%%
  %12 citations counted in INSPIRE as of 02 Mar 2020

  %\cite{Valderrama:2018bmv}
\bibitem{Valderrama:2018bmv}
  M.~P.~Valderrama,
  %``$\Omega(2012)$ as a hadronic molecule,''
  Phys.\ Rev.\ D {\bf 98}, 054009 (2018).
%  doi:10.1103/PhysRevD.98.054009
%  [arXiv:1807.00718 [hep-ph]].
  %%CITATION = doi:10.1103/PhysRevD.98.054009;%%
  %10 citations counted in INSPIRE as of 02 Mar 2020

%\cite{Huang:2018wth}
\bibitem{Huang:2018wth}
  Y.~Huang, M.~Z.~Liu, J.~X.~Lu, J.~J.~Xie and L.~S.~Geng,
  %``Strong decay modes $\bar{K}\Xi$ and $\bar{K}\Xi\pi$ of the $\Omega(2012)$ in the $\bar{K}\Xi(1530)$ and $\eta\Omega$ molecular scenario,''
  Phys.\ Rev.\ D {\bf 98}, 076012 (2018).
%  doi:10.1103/PhysRevD.98.076012
%  [arXiv:1807.06485 [hep-ph]].
  %%CITATION = doi:10.1103/PhysRevD.98.076012;%%
  %8 citations counted in INSPIRE as of 19 Feb 2020

%\cite{Polyakov:2018mow}
\bibitem{Polyakov:2018mow}
  M.~V.~Polyakov, H.~D.~Son, B.~D.~Sun and A.~Tandogan,
  %``¦¸(2012) through the looking glass of flavour SU (3),''
  Phys.\ Lett.\ B {\bf 792}, 315 (2019).
%  doi:10.1016/j.physletb.2019.03.054
%  [arXiv:1806.04427 [hep-ph]].
  %%CITATION = doi:10.1016/j.physletb.2019.03.054;%%
  %14 citations counted in INSPIRE as of 01 Jul 2020

%\cite{Lin:2019tex}
\bibitem{Lin:2019tex}
  Y.~H.~Lin, F.~Wang and B.~S.~Zou,
  %``Reanalysis of the newly observed $\Omega^*$ state in hadronic molecule model,''
  arXiv:1910.13919 [hep-ph].
  %%CITATION = ARXIV:1910.13919;%%
  %3 citations counted in INSPIRE as of 01 Jul 2020

 %\cite{Jia:2019eav}
\bibitem{Jia:2019eav}
  S.~Jia {\it et al.} [Belle Collaboration],
  %``Search for $\Omega(2012)\to K\Xi(1530) \to K\pi\Xi$ at Belle,''
  Phys.\ Rev.\ D {\bf 100}, 032006 (2019).
%  doi:10.1103/PhysRevD.100.032006
%  [arXiv:1906.00194 [hep-ex]].
  %%CITATION = doi:10.1103/PhysRevD.100.032006;%%
  %3 citations counted in INSPIRE as of 19 Feb 2020

%\cite{Lu:2020ste}
\bibitem{Lu:2020ste}
  J.~X.~Lu, C.~H.~Zeng, E.~Wang, J.~J.~Xie and L.~S.~Geng,
  %``Revisiting the $\Omega(2012)$ as a hadronic molecule and its strong decays,''
  Eur.\ Phys.\ J.\ C {\bf 80}, 361 (2020).
%  doi:10.1140/epjc/s10052-020-7944-4
%  [arXiv:2003.07588 [hep-ph]].
  %%CITATION = doi:10.1140/epjc/s10052-020-7944-4;%%
  %1 citations counted in INSPIRE as of 18 Jun 2020

%\cite{Ikeno:2020vqv}
\bibitem{Ikeno:2020vqv}
  N.~Ikeno, G.~Toledo and E.~Oset,
  %``Molecular picture for the $\Omega(2012)$  revisited,''
  Phys.\ Rev.\ D {\bf 101}, 094016 (2020).
%  doi:10.1103/PhysRevD.101.094016
%  [arXiv:2003.07580 [hep-ph]].
  %%CITATION = doi:10.1103/PhysRevD.101.094016;%%

%\cite{Oset:2016lyh}
\bibitem{Oset:2016lyh}
  E.~Oset {\it et al.},
%  ``Weak decays of heavy hadrons into dynamically generated resonances,''
  Int.\ J.\ Mod.\ Phys.\ E {\bf 25}, 1630001 (2016).
%  doi:10.1142/S0218301316300010
  %%CITATION = doi:10.1142/S0218301316300010;%%
  %4 citations counted in INSPIRE as of 25 Mar 2016

%\cite{Hyodo:2011js}
\bibitem{Hyodo:2011js}
  T.~Hyodo and M.~Oka,
%  ``Determination of the $\pi \Sigma$ scattering lengths from the weak decays of $\Lambda_c$,''
  Phys.\ Rev.\ C {\bf 84}, 035201 (2011).
%  [arXiv:1105.5494 [nucl-th]].
  %%CITATION = doi:10.1103/PhysRevC.84.035201;%%
  %10 c

%\cite{Miyahara:2015cja}
\bibitem{Miyahara:2015cja}
  K.~Miyahara, T.~Hyodo and E.~Oset,
%  ``Weak decay of $\Lambda_{c}^+$ for the study of $\Lambda$(1405) and $\Lambda$(1670),''
  Phys.\ Rev.\ C {\bf 92}, 055204 (2015).
%  [arXiv:1508.04882 [nucl-th]].
  %%CITATION = doi:10.1103/PhysRevC.92.055204;%%
  %9 citations counted in INSPIRE as of 29 janv. 2016


%\cite{Xie:2016evi}
\bibitem{Xie:2016evi}
  J.~J.~Xie and L.~S.~Geng,
  %``The $a_0(980)$ and $\Lambda(1670)$ in the $\Lambda^+_c \to \pi^+ \eta \Lambda$ decay,''
  Eur.\ Phys.\ J.\ C {\bf 76}, 496 (2016).
%  doi:10.1140/epjc/s10052-016-4342-z
%  [arXiv:1604.02756 [nucl-th]].
  %%CITATION = doi:10.1140/epjc/s10052-016-4342-z;%%
  %10 citations counted in INSPIRE as of 19 Jun 2020

%\cite{Xie:2017xwx}
\bibitem{Xie:2017xwx}
  J.~J.~Xie and L.~S.~Geng,
  %``$\Sigma^*_{1/2^-}(1380)$ in the $\Lambda^+_c \to \eta \pi^+ \Lambda$ decay,''
  Phys.\ Rev.\ D {\bf 95}, 074024 (2017).
%  doi:10.1103/PhysRevD.95.074024
%  [arXiv:1703.09502 [hep-ph]].
  %%CITATION = doi:10.1103/PhysRevD.95.074024;%%
  %10 citations counted in INSPIRE as of 19 Jun 2020

%\cite{Xie:2017erh}
\bibitem{Xie:2017erh}
  J.~J.~Xie and L.~S.~Geng,
  %``Role of the $N^*(1535)$ in the $\Lambda^+_c \to \bar{K}^0 \eta p$ decay,''
  Phys.\ Rev.\ D {\bf 96}, 054009 (2017).
%  doi:10.1103/PhysRevD.96.054009
%  [arXiv:1704.05714 [hep-ph]].
  %%CITATION = doi:10.1103/PhysRevD.96.054009;%%
  %4 citations counted in INSPIRE as of 19 Jun 2020

%\cite{Xie:2017mbe}
\bibitem{Xie:2017mbe}
  J.~J.~Xie and F.~K.~Guo,
  %``Triangular singularity and a possible $\phi p$ resonance in the $\Lambda^+_c \to \pi^0 \phi p$ decay,''
  Phys.\ Lett.\ B {\bf 774}, 108 (2017).
%  doi:10.1016/j.physletb.2017.09.060
%  [arXiv:1709.01416 [hep-ph]].
  %%CITATION = doi:10.1016/j.physletb.2017.09.060;%%
  %20 citations counted in INSPIRE as of 19 Jun 2020

%\cite{Liu:2019dqc}
\bibitem{Liu:2019dqc}
  X.~H.~Liu, G.~Li, J.~J.~Xie and Q.~Zhao,
  %``Visible narrow cusp structure in $\Lambda_c^+\to p K^- \pi^+$ enhanced by triangle singularity,''
  Phys.\ Rev.\ D {\bf 100}, 054006 (2019).
%  doi:10.1103/PhysRevD.100.054006
%  [arXiv:1906.07942 [hep-ph]].
  %%CITATION = doi:10.1103/PhysRevD.100.054006;%%
  %8 citations counted in INSPIRE as of 28 Jun 2020

%\cite{Sumihama:2018moz}
\bibitem{Sumihama:2018moz}
  M.~Sumihama {\it et al.} [Belle Collaboration],
  %``Observation of $\Xi(1620)^0$ and evidence for $\Xi(1690)^0$ in $\Xi_c^+ \rightarrow \Xi^-\pi^+\pi^+$ decays,''
  Phys.\ Rev.\ Lett.\  {\bf 122}, 072501 (2019).
%  doi:10.1103/PhysRevLett.122.072501
%  [arXiv:1810.06181 [hep-ex]].
  %%CITATION = doi:10.1103/PhysRevLett.122.072501;%%
  %6 citations counted in INSPIRE as of 01 Jul 2020

%\cite{Lu:2016ogy}
\bibitem{Lu:2016ogy}
  C.~D.~L\"{u}, W.~Wang and F.~S.~Yu,
  %``Test flavor SU(3) symmetry in exclusive $\Lambda_c$ decays,''
  Phys.\ Rev.\ D {\bf 93}, 056008 (2016).
%  doi:10.1103/PhysRevD.93.056008
%  [arXiv:1601.04241 [hep-ph]].
  %%CITATION = doi:10.1103/PhysRevD.93.056008;%%
  %59 citations counted in INSPIRE as of 19 Jun 2020

%\cite{Oset:1997it}
\bibitem{Oset:1997it}
  E.~Oset and A.~Ramos,
  %``Nonperturbative chiral approach to s wave anti-K N interactions,''
  Nucl.\ Phys.\ A {\bf 635}, 99 (1998).
%  doi:10.1016/S0375-9474(98)00170-5
%  [nucl-th/9711022].
  %%CITATION = doi:10.1016/S0375-9474(98)00170-5;%%
  %748 citations counted in INSPIRE as of 28 Jun 2020

%\cite{Liang:2014tia}
\bibitem{Liang:2014tia}
  W.~H.~Liang and E.~Oset,
  %``$B^0$ and $B^0_s$ decays into $J/\psi$ $f_0(980)$ and $J/\psi$ $f_0(500)$ and the nature of the scalar resonances,''
  Phys.\ Lett.\ B {\bf 737}, 70 (2014).
%  doi:10.1016/j.physletb.2014.08.030
%  [arXiv:1406.7228 [hep-ph]].
  %%CITATION = doi:10.1016/j.physletb.2014.08.030;%%
  %78 citations counted in INSPIRE as of 19 Jun 2020

%\cite{Xie:2018rqv}
\bibitem{Xie:2018rqv}
  J.~J.~Xie and G.~Li,
  %``The decays of $\bar{B}^0$ , $\bar{B}^0_s$ and $B^-$ into $\eta _c$ plus a scalar or vector meson,''
  Eur.\ Phys.\ J.\ C {\bf 78}, 861 (2018).
%  doi:10.1140/epjc/s10052-018-6353-4
%  [arXiv:1801.00536 [hep-ph]].
  %%CITATION = doi:10.1140/epjc/s10052-018-6353-4;%%
  %2 citations counted in INSPIRE as of 19 Jun 2020

%\cite{Xie:2018gbi}
\bibitem{Xie:2018gbi}
  J.~J.~Xie and E.~Oset,
  %``Search for the $\Sigma^*$ state in $\Lambda^+_c \to \pi^+ \pi^0 \pi^-\Sigma^+$ decay by triangle singularity,''
  Phys.\ Lett.\ B {\bf 792}, 450 (2019).
%  doi:10.1016/j.physletb.2019.04.011
%  [arXiv:1811.07247 [hep-ph]].
  %%CITATION = doi:10.1016/j.physletb.2019.04.011;%%
  %4 citations counted in INSPIRE as of 19 Jun 2020

%\cite{Jing:2020tth}
\bibitem{Jing:2020tth}
  H.~J.~Jing, C.~W.~Shen and F.~K.~Guo,
  %``Graphic Method for Phase Space Calculation,''
  arXiv:2005.01942 [hep-ph].
  %%CITATION = ARXIV:2005.01942;%%


\end{thebibliography}
\end{document}